\documentclass[preprint,showkeys,aip]{revtex4-1}
\usepackage{amsmath}
\usepackage{latexsym}
\usepackage{amssymb}
\usepackage{graphics,epstopdf}
\usepackage{amsmath,amssymb,amsthm}
\usepackage{graphicx}
\usepackage[colorlinks=true, citecolor=blue, urlcolor=blue ]{hyperref}
\usepackage{cleveref}

\begin{document}
	\title{On the  entanglement induced by the deformation of phase-space }
	\author{Shilpa Nandi}
	\email{nandishilpa801325@gmail.com}
	\affiliation{Department of Physics, Brahmananda Keshab Chandra College, 111/2 B. T. Road, Kolkata, India-700108}
	\author{Shatarupa Maity}
	\email{shatarupamaity2000@gmail.com}
	\affiliation{Department of Physics, Brahmananda Keshab Chandra College, 111/2 B. T. Road, Kolkata, India-700108}
	\author{Pinaki Patra}
	\thanks{Corresponding author}
	\email{monk.ju@gmail.com}
	\affiliation{Department of Physics, Brahmananda Keshab Chandra College, 111/2 B. T. Road, Kolkata, India-700108}

	\date{\today}

	\begin{abstract}
Most quantum gravity theories propose that the fundamental concept of space-time is mostly compatible with quantum theory in noncommutative (NC) space. In the present paper, we revisit the notion of entanglement induced by NC deformations of phase space.  The positive partial transpose (PPT) criterion for separability of bipartite Gaussian states is extended to a general class of Bopp's shift.  In particular, we have considered both the position-position and momentum-momentum noncommutativity, with deformation parameters $\theta$ and $\eta$, respectively. It turns out that $\theta$ and $\eta$ induce the entanglement. We have directly applied the formalism for an anisotropic two-dimensional harmonic oscillator. Peres-Horodecki separability condition leads to a constraint equation for the parameter values of the oscillator in NC space. It turns out that the bipartite Gaussian state is almost always entangled in deformed space. To implement the theoretical idea, we provide an outline for a gedankenexperiment to identify the signature of phase-space noncommutativity, i.e., quantum gravity. In particular, the gedankenexperiment is devised to test the separability of supposedly separable Gaussian states in the usual commutative space, through the covariance matrix, which is constructed via measured output photocurrents after interaction of input Gaussian states and reference states. If the experiment shows that the supposedly separable states are actually entangled, then the entanglement is created through the intermediate background noncommutative space, which is a signature of the quantum nature of gravity. 
	\end{abstract}

		
	\maketitle
	
	\section{Introduction}
	Quantum entanglement (verschr\"{a}nkung) implies the existence of global states of a composite system which can not be written as a product of the states of individual
subsystems \cite{entanglement1}. It underlines the intrinsic order of statistical correlations between subsystems of a compound quantum system \cite{entanglement2}. The insight into quantum entanglement leads to an understanding of a diverse range of phenomena in which correlations are important \cite{decoherence}. For instance, the understanding of quantum optics, phase transition in condensed matter systems, black hole physics, and the relation between the geometrical structure of the dual spacetime and entanglement structure of the conformal field theory have become topical issues \cite{entanglement3,entanglement4,entanglement5,entanglement6}. One of the key goals of modern physics is to understand the physical origin(s) of entanglement, which is still obscure \cite{entorigin}. However, one thing is clear- entanglement is a purely quantum phenomenon. Entanglement can not be generated through any classical mediator \cite{classmed1}. 
	\\
    Besides quantum mechanics (QM), we have another cornerstone of modern physics, namely Gravity, which describes the dynamics of spacetime itself \cite{rovelli1}. There is no single reproducible reported violation of QM and gravity to date. We expect that QM and gravity are subsets of some unified theory, namely quantum gravity (QG). There are many elegant, consistent theoretical proposals of QG, namely string theory, loop quantum gravity, M-theory, twistor theory, and so on \cite{quantumgravity1,quantumgravity2,quantumgravity3,quantumgravity4}. However, the lack of experimental confirmation of any of the proposals of QG raises serious doubts about whether gravity falls under the premise of quantum theory at all \cite{hanif1}. Most probably, the lack of successful experiments for QG is not due to the drawback of the proposed theories. Rather, it is due to the limitation of present-day experimental capacity. By way of comparison, the LHC was designed to run at a maximum collision energy of $14$  TeV \cite{lhc}, whereas most of the theories of quantum gravity appear to predict departures from classical gravity only at energy scales on the order of $10^{19}$ GeV \cite{qg1,qg2}. We hope that the high-energy scattering experiment may lead us to the feasibility of probing such high-energy scales in the future. However, colliders are costly. Naturally, alternative directions toward the exploration of the quantum signature of gravity are always fascinating. For instance,  proposals  toward low-energy laboratory tests that can witness the nonclassicality of the gravitational field, with the help of conventional interferometric techniques, are available in the literature \cite{expt1,expt2,expt3}.
    \\
   In the line of investigation towards the proposals of experimental test for QG, we utilize the consensus among most of the theoretical proposals that the space-time structure is deformed in such a manner that the usual notion of commutative space is ceased at the energy scales in which it is predicted to show departures from classical gravity \cite{qg4,qg5}. In particular,  the fundamental concept of space-time is mostly compatible with quantum theory in noncommutative (NC) space \cite{qg6,qg7}. In other words, at very high energies, a common expectation is that space-time may not retain its smooth continuous structure at very short distances \cite{qg8,ncs1}.  Moyal deformation of ordinary space-time is an example of a specific algebraic realization of this expectation \cite{Moyal1,ncs3}. 
   It seems likely that any modification to the symplectic structure of phase-space, as brought about by position-position noncommutativity and momentum-momentum noncommutativity at a fundamental scale, could impact the correlations of coordinate and momentum degrees of freedom \cite{nandi1,Moyal2,ncssim1,Gouba1,NCent1}. In particular, we expect a deep-rooted connection between noncommutative dynamics and the phenomenon of entanglement of quantum states \cite{NCent2,NCent3,NCent4}. 
	\\
	In the present paper, we investigate whether some optomechanical scheme can predict the signature of noncommutativity of space-time. The scheme relies on the basic premise that if no other effect is present, and we observe that a supposedly separable state shows an entanglement property, then this departure from the prediction of usual quantum mechanics is due to a mediated gravitational field. Since entanglement is a purely quantum phenomenon, this implies the intermediate background space is quantum in nature. 
	The proposed scheme is based on a massive noncommutative oscillator to interact with reference optical fields in an optomechanical cavity by utilizing the principle of radiation pressure interaction \cite{expt2,expt3,marleto1,sibasish1}. We measure the output photocurrent and estimate the expectation values of quadratures for various phases of input optical modes. The data of photocurrents provide the covariance matrix for the system \cite{sibasish1}.   Signature of deformation of the commutation relation of phase-space operators will be carried in the data of photocurrent. For Gaussian states, the covariance matrix provides  the signature of entanglement through a sufficient condition, namely, Peres-Horodecki separability criteria \cite{gaussian1,ppt1}. 
	From a viewpoint of constructor-theoretic principles, the proposed optomechanical schemes are promising. In particular, if we observe the entanglement effects
	in the measurement of the properties of two quantum masses that interact with each other
	through gravity only, then we can conclude that the mediator (gravity) has to have some
	quantum features \cite{marleto1}. It doesn’t matter in what way gravity is quantum - whether it is loop quantum gravity or string theory or something else - but it has to be a quantum theory \cite{marleto1,NCent2}.
	 \\ 
	 On the other hand, Gaussian states (GS) play a key role in quantum optics as all processes generated by Hamiltonians up to second order in the field operators (i.e., linear optics and quadrature squeezing) preserve Gaussianity \cite{optics1}.  The simplicity of GS is that it can be completely determined by its covariance matrix (CVM) $\Sigma$ and first-moment vector \cite{Gaussianref1,Gaussianref2}. The first-moment vector can be set to zero for all practical purposes with coordinate shifting. For that reason, we choose the Gaussian state, in particular, the ground state of the harmonic oscillator in NCS, for the study of the present paper.
	 For a working model, we choose the quantum gravity models, in which we consider noncommutative spatial operators ($[x,y]=i\theta$) and momentum operators ($[p_x,p_y]=i\eta$). A quantum system under the phase-space deformation can be mapped into an equivalent system in usual quantum mechanics through Bopp's shift, which is a nonsymplectic transformation in phase space.  
	 It is worth noting that the NC-space parameter-dependent entanglement has been studied in the literature, and it is fairly well known \cite{NCent1,NCent2,NCent3,NCent4}.  In the present paper, we start with a bipartite  (supposedly separable) Gaussian state shared by two observers, Alice (A) and Bob (B). A and B are unaware of whether the background space is noncommutative or not. After interaction of the input state with the reference optical beams, the output photocurrents are measured, and they construct the covariance. They can utilize the separability criteria for a bipartite Gaussian state and infer about the entanglement. If it is found that the states are entangled, then it must have been generated through an intermediate gravitational field. Thus, it is a signature of the quantum nature of gravity.
	 \\
The organization of the paper follows. At first, we discuss the entanglement generated by the noncommutative deformation of phase space for an anisotropic harmonic oscillator. After that, we explicitly provide the generation of entanglement for a generic bipartite Gaussian state through Bopp's shift.  Then we have outlined a gedankenexperiment for experimental implementation of the theoretical results corresponding to the signature of noncommutativity of space-time, i.e., quantum gravity. Finally, we discuss our results. 
\section{Entanglement generation by non-commutative deformations}
Suppose Alice and Bob are performing measurements along the $\tilde{x}_1$ and $\tilde{x}_2$ axis of a NC- space, respectively, on an anisotropic oscillator described by the Hamiltonian 
\begin{equation}\label{NCShamiltonian}
	\hat{H}_{nc}=\frac{1}{2} \hat{\tilde{X}}^T \mathcal{H}_{nc}\hat{\tilde{X}},\; \mbox{with}\; \hat{\tilde{X}}= (\hat{\tilde{X}}_1,\hat{\tilde{X}}_2, \hat{\tilde{X}}_3, \hat{\tilde{X}}_4)^T= (\hat{\tilde{x}}_1,\hat{\tilde{p}}_1, \hat{\tilde{x}}_2, \hat{\tilde{p}}_2)^T.
\end{equation}
Here,  $X^T$ stands for matrix transposition of $X$. The anisotropic oscillator in the NC-space is characterized by   the mass $M=(m_1,m_2)$, and the frequency $\tilde{\omega}=(\tilde{\omega}_1,\tilde{\omega}_2)$ through
\begin{eqnarray}
\mathcal{H}_{nc}= \left(\begin{array}{cc}
	\mathcal{H}_{nc}^{(1)} & 0 \\
	0 & \mathcal{H}_{nc}^{(2)}
\end{array}\right),\;
\mbox{with}\;
\mathcal{H}_{nc}^{(j)}=\mbox{Diag}(m_j\tilde{\omega}_j^2,1/m_j),\; j=1,2.
\end{eqnarray}
We consider the fundamental commutation relations  in NC-space  as 
\begin{equation}\label{NCspacecommutation}
	[	\hat{\tilde{X}}_{\alpha}, \hat{\tilde{X}}_{\beta}]= i\hbar_e \tilde{J}_{\alpha\beta}= - (\tilde{\Sigma}_y)_{\alpha\beta},
\end{equation}
where the deformed symplectic matrix $\tilde{J}$ and the effective Planck constant $\hbar_e$ reads
\begin{eqnarray}
	\tilde{J} = \left(\begin{array}{cc}
		J_2 &  \frac{1}{\hbar_e}\Pi_{\theta\eta}\\
		-\frac{1}{\hbar_e}\Pi_{\theta\eta} &  J_2
	\end{array}\right),\; \mbox{with}\; 
	\Pi_{\theta\eta}= \left(\begin{array}{cc}
		\theta &  0\\
		0 & \eta
	\end{array}\right),\; 
	\hbar_{e}=\hbar(1+\frac{\theta\eta}{4\hbar^2}).
\end{eqnarray}
Here $\theta$ and $\eta$ are the position position NC-parameter and momentum-momentum NC-parameter, respectively. $\hbar$ is the Planck constant, and $J_2$ is the usual symplectic matrix, which encodes the commutation relations in the usual commutative space 	$\hat{X}=(\hat{X}_1, \hat{X}_2,\hat{X}_3,\hat{X}_4)^T=(\hat{x}_1, \hat{p}_1,\hat{x}_2,\hat{p}_2)^T$ as follows.
\begin{eqnarray}\label{Cspacecommutation}
	[	\hat{X}_{\alpha}, \hat{X}_{\beta}]=i\hbar J_{\alpha\beta}= -\hbar(\Sigma_y)_{\alpha\beta}, \;
	\mbox{with}\;
	J=\mbox{diag}(J_2,J_2),\\ \Sigma_j=\mbox{diag}(\sigma_j, \sigma_j),\; \mbox{for}\; j=x,y,z.\nonumber
\end{eqnarray}
Here the symplectic matrix $J_2$ and Pauli matrices are represented by
\begin{eqnarray}
	\sigma_x =\left(\begin{array}{cc}
		0 & 1\\
		1& 0
	\end{array}\right),\;
	\sigma_y =\left(\begin{array}{cc}
		0 & -i\\
		i& 0
	\end{array}\right),\;
	\sigma_z =\left(\begin{array}{cc}
		1 & 0\\
		0& -1
	\end{array}\right),\;
	J_2 = \left(\begin{array}{cc}
		0 & 1\\
		-1& 0
	\end{array}\right).
\end{eqnarray}

The NC-space co-ordinates ($\hat{\tilde{X}}$) are connected to the  commutative space co-ordinates ($\hat{X}$) through the Darboux transformation ($\Upsilon_D$) given by the Bopp's shift
\begin{eqnarray}\label{cncconnection}
	\hat{\tilde{X}}=\Upsilon_D \hat{X},\;\;
	\mbox{with}\;
	\Upsilon_D =\left(\begin{array}{cc}
		\mathbb{I}_2 & -\frac{1}{2\hbar}\Pi_{\theta\eta}J_2 \\
		\frac{1}{2\hbar}\Pi_{\theta\eta}J_2 & \mathbb{I}_2 
	\end{array}\right).
\end{eqnarray}
The notation $\mathbb{I}_n$ stands for $n\times n$ identity matrix.
Here our concern is to have a valid Darboux transformation. Thus we restrict the parameter choice such that $\frac{\theta\eta}{4\hbar^2} <1$. In other words, we restrict the determinant of $\Upsilon_D $ is nonzero ($\Delta_{\Upsilon_D}\neq 0$); i.e., $\Upsilon_D \in GL(4, R)$.
$J$ is connected with the deformed symplectic matrix $\tilde{J}$ through 
\begin{equation}\label{JncJcconnection}
	\hbar_e \tilde{J} = \hbar \Upsilon_D J\Upsilon_D^T .
\end{equation}
Since the quantum mechanical formalism are well established in commutative space, it is customary to convert the NC-space system in the usual commutative space system through ~\eqref{cncconnection} for computational purpose.
Using ~\eqref{cncconnection}, one can see that the NC-space Hamiltonian ~\eqref{NCShamiltonian} is equivalent to the usual commutative space Hamiltonian 
\begin{eqnarray}\label{Hamiltonianmatrix}
	\hat{H}=\frac{1}{2} \hat{X}^T \mathcal{H}\hat{X},\; \mbox{with}\; \mathcal{H}=\Upsilon_D^T \mathcal{H}_{nc}\Upsilon_D.
\end{eqnarray}
Explicitly written
\begin{eqnarray}
	\mathcal{H}=\left(\begin{array}{cc}
\mathcal{H}_1 & \mathcal{H}_{12}\\
\mathcal{H}_{12}^T & \mathcal{H}_2
	\end{array}\right),\;
\mbox{with}\; \mathcal{H}_{12}=\left(\begin{array}{cc}
	0 & -2\nu_2 \\
	2\nu_1 & 0
\end{array}\right),\; \mathcal{H}_j=\mbox{Diag}(\mu_j\omega_j^2,1/\mu_j); j=1,2
.
\end{eqnarray}
Explicit forms of the  parameters are given by
\begin{eqnarray}
	\frac{1}{\mu_1} &=& \frac{1}{m_1}+ \frac{1}{4\hbar^2} m_2\tilde{\omega}_2^2\theta^2, \;
	\frac{1}{\mu_2} = \frac{1}{m_2}+ \frac{1}{4\hbar^2} m_1\tilde{\omega}_1^2\theta^2,\\
	\alpha_1&=& \mu_1\omega_1^2=m_1\tilde{\omega}_1^2 + \frac{\eta^2}{4\hbar^2 m_2} ,\;
	\alpha_2  = \mu_2\omega_2^2=m_2\tilde{\omega}_2^2 + \frac{\eta^2}{4\hbar^2 m_1},\\
	\nu_1 & =& \frac{1}{4m_1\hbar} (\eta + m_1 m_2\tilde{\omega}_2^2 \theta),\;
	\nu_2 = \frac{1}{4m_2\hbar} (\eta + m_1 m_2\tilde{\omega}_1^2 \theta).
\end{eqnarray}
Note that for isotropic oscillator $\mu_1=\mu_2,\; \alpha_1=\alpha_2,\; \nu_1=\nu_2$. For an isotropic oscillator, one can identify the off-diagonal terms constitute the angular momentum operator. \\
One can diagonalize the bilinear Hamiltonian ~\eqref{Hamiltonianmatrix} with a normal coordinate system, keeping the symplectic structure $Sp(4,\mathbb{R})$ intact. First we note that, the symplectic eigenvalues \cite{Williamson1,Williamson2} of $\mathcal{H}$ are the ordinary eigenvalues of $
	\mathcal{H}_J=J\mathcal{H}$.
Since $	\mathcal{H}_J$ is not symmetric, the left and right eigenvectors of $	\mathcal{H}_J$ are not the same. However, left and right eigenvalues are identical. The characteristic polynomial $	P_{\mathcal{H}_J}(\lambda)$ of $\mathcal{H}_J$ has four distinct purely imaginary roots for the  real parameters $\mu_j,\omega_j,\nu_j$:
\begin{equation}\label{lambdaformexplicit}
	\lambda \in \{\mp i\lambda_j, j=1,2\vert \lambda_1=\sqrt{(\Delta-D)/2}, \lambda_2=\sqrt{(\Delta+D)/2} \},
\end{equation}
where $\Delta=\Delta_{\mathcal{H}_1}+\Delta_{\mathcal{H}_2}+2\Delta_{\mathcal{H}_{12}}$, and the discriminant $D=\sqrt{\Delta^2-4\Delta_{\mathcal{H}_J}}$ is given by
\begin{eqnarray}\label{discriminant}
	D^2= (\omega_1^2-\omega_2^2)^2 + 16\nu_1\nu_2 (\omega_1-\omega_2)^2 + 16 \left(\sqrt{\frac{\mu_1}{\mu_2}}\omega_1\nu_1 + \sqrt{\frac{\mu_2}{\mu_1}}\omega_2\nu_2\right)^2\ge 0.
\end{eqnarray}
Here we have used the notation $\Delta_A=\mbox{Det}(A)$.
Note that $D=0$ only for parameter values $\omega_1=\omega_2,\; \nu_1=\nu_2=0$, which corresponds to the isotropic oscillator in commutative space. We shall consider $D>0$ for our present study.

If $\chi_{lj}$ is the left eigenvector corresponding to the eigenvalue $-i\lambda_j$ of $\mathcal{H}_J$, i.e., $\chi_{lj}\mathcal{H}_J=-i\lambda_j \chi_{lj}$, then the direct computation gives
\begin{eqnarray}
	\chi_{lj}=k_{j}(i \kappa_{j,1}, \kappa_{j,2}, \kappa_{j,3}, i\kappa_{j,4});\; j=1,2.
\end{eqnarray}
Here $k_{j}$ is the normalization constant, and the real parameters $\kappa_{j,\alpha}^{r} $ and $\kappa_{j,\alpha}$ reads
\begin{eqnarray}
	\kappa_{j,1} &=& -2\mu_1\lambda_j (\mu_1\nu_1\omega_1^2 + \mu_2\nu_2\omega_2^2),\;
	\kappa_{j,2} = 2(\mu_2\nu_2\omega_2^2 - 4\mu_1\nu_1^2 \nu_2 + \mu_1\nu_1\lambda_j^2), \label{kappaj12}\\
	\kappa_{j,3} &=& \mu_1(4\mu_1\nu_1^2\omega_1^2 - \mu_2 \omega_1^2 \omega_2^2 + \mu_2\omega_2^2 \lambda_j^2),\;
	\kappa_{j,4} = -\mu_1\lambda_j (\omega_1^2 + 4\nu_1\nu_2 -\lambda_j^2). \label{kappaj34}
\end{eqnarray}
The left eigenvector corresponding to the eigenvalue $i\lambda_j$ is given by $	\chi_{lj}^*$. The right eigenvector $\chi_{rj}$ corresponding to the eigenvalue $-i\lambda_j$   may be obtained through   $\chi_{rj}=-\Sigma_y \chi_{lj}^\dagger$. Normalization condition $\chi_{lj}\chi_{rj}=1$ yields
\begin{equation}\label{normalizationofchi}
	\vert k_{j}\vert= 1/\sqrt{2(\kappa_{j,3}\kappa_{j,4}- \kappa_{j,1} \kappa_{j,2})};\; j=1,2. 
\end{equation}
Similarity transformation which diagonalizes $\mathcal{H}_J$, i.e., 
\begin{equation}
	\mathcal{H}_{JD}=\mbox{diag}(-i\lambda_1,i\lambda_1,-i\lambda_2,-i\lambda_2)= Q^{-1}\mathcal{H}_J Q,
\end{equation}
is given by the following matrices
\begin{equation}
	Q=(\chi_{r1},\chi_{r1}^*,\chi_{r2},\chi_{r2}^*),\;	Q^{-1}=(\chi_{l1}^T,\chi_{l1}^{*T},\chi_{l2}^T,\chi_{l2}^{*T})^T.
\end{equation}
The diagonal representation of $\mathcal{H}_J$ enables to  define the normal co-ordinates through
\begin{equation}\label{a1a2}
	\hat{A}=(\hat{a}_1,\hat{a}_1^\dagger,\hat{a}_2,\hat{a}_2^\dagger)^T=\frac{1}{\sqrt{\hbar}}Q^{-1}\hat{X}.
\end{equation}
Using the normalization condition ~\eqref{normalizationofchi}, it follows 
\begin{equation}\label{algebraaadagger}
	[\hat{a}_1,\hat{a}_1^\dagger]=	[\hat{a}_2,\hat{a}_2^\dagger]=1.
\end{equation}
On the other hand, the orthogonality condition $\chi_{l1}\chi_{r2}=\chi_{l2}\chi_{r1}=0$ is equivalent to
\begin{equation}\label{algebraa1a2}
	[\hat{a}_1,\hat{a}_2]=0.
\end{equation}
The algebra ~\eqref{algebraaadagger} and ~\eqref{algebraa1a2} confirm that $\hat{a}_j$ and $\hat{a}_j^\dagger$ are annihilation and creation operators.
The ground state of the system thus satisfy the property
\begin{equation}\label{groundstatedefnforanisotropic}
	\hat{a}_1\vert 0,0\rangle =\hat{a}_2\vert 0,0\rangle =0.
\end{equation}
In position representation ($\psi_{0,0}(x_1,x_2)=\langle x_1,x_2\vert 0,0\rangle$), the equation ~\eqref{groundstatedefnforanisotropic} reads
\begin{equation}\label{groundstateforanisotropic}
	(U_xx-i\hbar U_p\partial_{x})\psi_{0,0}(x_1,x_2)=0,
\end{equation}
where
\begin{eqnarray}
	x=\left(\begin{array}{c}
		x_1 \\
		x_2
	\end{array}\right),\; 
	\partial_x=\left(\begin{array}{c}
		\frac{\partial}{\partial x_1} \\
		\frac{\partial}{\partial x_2}
	\end{array}\right),\;
	U_x=\left(\begin{array}{cc}
		i\kappa_{11} & \kappa_{31}\\
		i\kappa_{12} & \kappa_{32}
	\end{array}\right),\;
	U_p=\left(\begin{array}{cc}
		\kappa_{21} & i\kappa_{41}\\
		\kappa_{22} & i\kappa_{42}
	\end{array}\right).
\end{eqnarray}
We take the following ansatz for the solution of ~\eqref{groundstateforanisotropic}.
\begin{equation}\label{ansatzpsi00anisotropic}
	\psi_{0,0}(x_1,x_2)= \mathcal{N}_0 e^{-\frac{1}{2}x^T\Lambda_{ancs}x}.
\end{equation}
$\mathcal{N}_0$ being the normalization condition and
\begin{eqnarray}
	\Lambda_{ancs}=\left(\begin{array}{cc}
		\Lambda_{11} & \Lambda_{12}\\
		\Lambda_{12} & \Lambda_{22}
	\end{array}\right).
\end{eqnarray}
. Using ~\eqref{ansatzpsi00anisotropic} in ~\eqref{groundstateforanisotropic} we get
\begin{equation}
	\Lambda_{ancs}=\frac{i}{\hbar} U_p^{-1}U_x.
\end{equation}

Explitly written,
\begin{eqnarray}
	\Lambda_{11} &=& \frac{\kappa_{41}\kappa_{12}-\kappa_{42}\kappa_{11}}{\hbar(\kappa_{21}\kappa_{42}-\kappa_{22}\kappa_{41})} =\Lambda_{11r},\;\;
	\Lambda_{22} = \frac{\kappa_{21}\kappa_{32}-\kappa_{22}\kappa_{31}}{\hbar(\kappa_{21}\kappa_{42}-\kappa_{22}\kappa_{41})}=\Lambda_{22r}, \label{Lambda11form}\\
	\Lambda_{12} &=& \frac{i(\kappa_{42}\kappa_{31}-\kappa_{41}\kappa_{32})}{\hbar(\kappa_{21}\kappa_{42}-\kappa_{22}\kappa_{41})}=\frac{i(\kappa_{12}\kappa_{21}-\kappa_{22}\kappa_{11})}{\hbar(\kappa_{21}\kappa_{42}-\kappa_{22}\kappa_{41})}\label{Lambda12form}=i\Lambda_{12c}.
\end{eqnarray}
In particular, $\Lambda_{11}$ and $\Lambda_{22}$ are real, whereas $\Lambda_{12}$ is purely imaginary. 
\subsection{Noise matrix and separability criterion}
Wigner distribution ($W$) and the density operator $(\hat{\rho})$ are related
through the definition
\begin{equation}
	W(X_c) =\frac{1}{\pi^2\hbar^2}\int d^2 x' \langle x-x'\vert \hat{\rho} \vert x+x'\rangle e^{2ix'.p/\hbar},
\end{equation}
where $X_c=(X_1,X_2,X_3,X_4)=(x_1,p_1,x_2,p_2)^T$ is the classical co-ordinate  vector. We define 
\begin{eqnarray}
	\Delta \hat{X}_\alpha &=& \hat{X}_\alpha -\langle \hat{X}_\alpha\rangle_\rho,  \mbox{with}\; \langle \hat{X}_\alpha\rangle_\rho = \mbox{Tr}(\hat{X}_\alpha\hat{\rho}); \;\alpha=1,2,3,4.\\
	\Delta X_\alpha &=& X_\alpha -\langle X_\alpha\rangle_W,\; \mbox{with}\; \langle X_\alpha\rangle_W = \int   W(X_c)X_\alpha d^4X_c .
\end{eqnarray}
The four components $\Delta \hat{X}$ satisfy the same commutation relations as $\hat{X}$. Moreover, the phase-space average $\langle X_\alpha\rangle_W$ with respect to the Wigner distribution $W$ is equal to the average $\langle \hat{X}_\alpha\rangle_\rho$ with respect to the density operator $\hat{\rho}$. We define the covariance matrix $\mathcal{V}_c$ through the matrix elements
\begin{equation}
	\mathcal{V}_{\alpha\beta}=\frac{1}{2} \langle \{\Delta \hat{X}_\alpha, \Delta \hat{X}_\beta\}\rangle =\mbox{Tr}(\frac{1}{2}\{\Delta \hat{X}_\alpha, \Delta \hat{X}_\beta\}\hat{\rho})=\int \Delta X_\alpha \Delta X_\beta W(X_c)d^4X_c.
\end{equation}
Since $\mathcal{V}_c$ is symmetric, we can  write it in the block form
\begin{eqnarray}\label{cspacecovariancematrix}
	\mathcal{V}_c=\left(\begin{array}{cc}
		V_{11} & V_{12}\\
		V_{12}^T & V_{22}
	\end{array}\right).
\end{eqnarray}
Using the connection ~\eqref{cncconnection} between the commutative space and NC-space co-ordinates, one can see that the NC-space covariance matrix $\mathcal{V}_{nc}$ is given by
\begin{equation}\label{VncVcconnection}
	\tilde{\mathcal{V}}_{nc}= \Upsilon_D  \mathcal{V}_c\Upsilon_D^T .
\end{equation}
The fundamental commutation relations ~\eqref{Cspacecommutation} of commutative space implies that a bonafide covariance matrix must satisfy the Robertson-Schr\"{o}dinger uncertainty principle (RSUP)
\begin{equation}\label{RSUPc}
	\mathcal{V}_c+\frac{i}{2}\hbar J \ge 0 .
\end{equation}
In NC-space, the equivalent statement for the RSUP reads
\begin{equation}\label{RSUPnc}
	\tilde{\mathcal{V}}_{nc} + \frac{i}{2} \hbar_e \tilde{J} \ge 0.
\end{equation}
A generic local transformation $S_1 \bigoplus S_2$, acts on $\mathcal{V}_c$ as
\begin{equation}
	V_{jj}\to S_j V_{jj} S_j^T ,\; 	V_{12}\to S_1 V_{12}S_2^T;\; \mbox{with}\; S_j \in Sp(2,\mathbb{R}),\;  j=1,2.
\end{equation}
One can identify that following four quantities are local  invariant with respect to transformation belonging to the $Sp(2,\mathbb{R})\bigotimes Sp(2,\mathbb{R}) \subset Sp(4,\mathbb{R})$.
\begin{eqnarray}
	\Delta_j=Det(V_{jj}),	\Delta_{12}=Det(V_{12}),  \Delta_{\mathcal{V}_c} = Det(\mathcal{V}_c),
	\tau_{v}=\mbox{Tr}(V_{11} J_2 V_{12} J_2 V_{22} J_2 V_{12}^T J_2).
\end{eqnarray}
Using Williamson's theorem \cite{Williamson1,Williamson2}, one can show that the RSUP ~\eqref{RSUPc} can be rewritten as $Sp(2,\mathbb{R})\bigotimes Sp(2,\mathbb{R})$ invariant statement
\begin{equation}\label{covariancersupc}
	\Delta_1 \Delta_2 + (\hbar^2/4- \Delta_{12})^2 -\tau_{v}\ge \hbar^2( \Delta_1+\Delta_2)/4.
\end{equation}
Under mirror reflection (Peres-Horodecki partial transpose) $\Delta_1$,  $\Delta_2$ and $\tau_v$ remain invariant; whereas $\Delta_{12}$ flips sign. Therefore, the requirement that the covariance matrix of a separable state has to obey the following necessary condition.
\begin{equation}\label{separabilityc}
	Ps=	\Delta_1 \Delta_2 + (\hbar^2/4-\vert \Delta_{12}\vert)^2 - \tau_{v}- \hbar^2( \Delta_1+\Delta_2)/4 \ge 0,
\end{equation}
which turns out to be sufficient for all bipartite-Gaussian state.\\
Now, for a generic ground state ~\eqref{groundstatedefnforanisotropic}, which has the form in position representation as
$\psi(x_1,x_2)=\mathcal{N}_0e^{-\frac{1}{2}x^T\Lambda x}$, corresponds to 
the Wigner distribution
\begin{equation}\label{WX}
	W(X)= \frac{1}{\pi^2\hbar^2} \exp\{ - x^T (\Lambda_r + \Lambda_c \Lambda_r^{-1}\Lambda_c^T)x - \frac{1}{\hbar^2} p^T\Lambda_r^{-1}p  -\frac{1}{\hbar} (x^T \Lambda_c \Lambda_r^{-1} p + p^T \Lambda_r^{-1}\Lambda_c^T x) \}.
\end{equation}
Here the following notation for the matrix $\Lambda$ and its elements $\Lambda_{jk}$ have been used.
\begin{eqnarray}
	\Lambda=[\Lambda_{jk}]_{j,k=1}^{2}= \Lambda_r +i\Lambda_c= [\Lambda_{jkr}]_{j,k=1}^{2}+i[\Lambda_{jkc}]_{j,k=1}^{2},
\end{eqnarray}
where $\Lambda_{jkr}=\Re(\Lambda_{jk}),\Lambda_{jkc}=\Im(\Lambda_{jk})$, and $x=(x_1,x_2)^T,\; p=(p_1,p_2)^T$.
The expectation values through the Wigner distribution $W(X)$, provide the covariance matrix $\mathcal{V}_{c}$ as
\begin{eqnarray}\label{vc}
	\mathcal{V}_{c}=  \frac{\hbar}{2}\left(\begin{array}{cc}
		\sigma_{11} & \sigma_{12}\\
		\sigma_{12}^T & \sigma_{22}
	\end{array}\right),
\end{eqnarray}
with
\begin{eqnarray}\label{formofsigmaij}
	\sigma_{11}=\left(\begin{array}{cc}
		\frac{1}{\hbar\Lambda_{11r}} & 0\\
		0 & \frac{\hbar\Delta_\Lambda}{\Lambda_{22r}}
	\end{array}\right),\;
	\sigma_{22}=\left(\begin{array}{cc}
		\frac{1}{\hbar\Lambda_{22r}} & 0\\
		0 & \frac{\hbar\Delta_\Lambda}{\Lambda_{11r}}
	\end{array}\right),\;
	\sigma_{12}=\left(\begin{array}{cc}
		0& -	\frac{\Lambda_{12c}}{\Lambda_{11r}} \\
		-\frac{\Lambda_{12c}}{\Lambda_{22r}} &0
	\end{array}\right),
\end{eqnarray}
where $\Delta_\Lambda= \Lambda_{11r}\Lambda_{22r}+\Lambda_{12c}^2$.
Using  ~\eqref{formofsigmaij} in the generalized Peres-Horodecki separability criterion ~\eqref{separabilityc} we get the following constraint on the parameters.
\begin{equation}
	-\Lambda_{11r} \Lambda_{22r} \Lambda_{12c}^2 \ge \Lambda_{12c}^2 \Lambda_{11r}\Lambda_{22r}\implies
	\Lambda_{11r}\Lambda_{22r}\Lambda_{12c}=0.
\end{equation}
Since $\Lambda_{11r}$ and $\Lambda_{22r}$ are nonzero, the separability of states implies $\Lambda_{12c}=0$, i.e.,
\begin{equation}\label{constrainteqn}
	(\lambda_2-\lambda_1)(\mu_2\nu_2\omega_2^2 -4\mu_1\nu_1^2\nu_2-\mu_1\nu_1\lambda_1\lambda_2)=0.
\end{equation}
However, according to ~\eqref{discriminant} and ~\eqref{lambdaformexplicit}, we consider $\lambda_1\neq \lambda_2$. Therefore, ~\eqref{constrainteqn} holds for
\begin{equation}
	(\mu_1^2\omega_1^2\nu_1^2-\mu_2^2\omega_2^2\nu_2^2)(\omega_2^2-4\mu_1\nu_1^2/\mu_2)=0.
\end{equation}
From the physical viewpoint, we choose $\theta,\eta \le \hbar$. In other words, $\theta\eta \neq 4\hbar^2$, which means $\omega_2^2 \neq 4\mu_1\nu_1^2/\mu_2$. Therefore, the only possibility for the separable states is satisfied by the constraint $\mu_1\nu_1\omega_1 =\mu_2\nu_2\omega_2$, which is equivalent to the following equation in terms of original parameters.
\begin{eqnarray}\label{sep1}
	(4\hbar^2/m_{12}+\tilde{\omega}_1^2\theta^2) (\eta/m_{12}+ \tilde{\omega}_2^2\theta)^2 (\eta^2/m_{12}+ 4\hbar^2\tilde{\omega}_1^2) 
	=  (4\hbar^2/m_{12}+\tilde{\omega}_2^2\theta^2) \nonumber \\
	(\eta/m_{12}+ \tilde{\omega}_1^2\theta)^2 (\eta^2/m_{12}+ 4\hbar^2\tilde{\omega}_2^2) ,\; \mbox{with}\; m_{12}=m_1.m_2.
\end{eqnarray}
The condition ~\eqref{sep1} is trivially satisfied for commutative space ($\theta,\eta\to 0$), and as well as for isotropic oscillator ($\tilde{\omega}_1=\tilde{\omega}_2$) in noncommutative (NC) space. Therefore, the entanglement between the coordinate degrees of freedom is not the sole property of the noncommutativity of space. It depends on both the NC-parameters ($\theta,\eta$) and the anisotropy of oscillator frequency ($\tilde{\omega}_1\neq\tilde{\omega}_2$). If we recall that the off-diagonal term for the isotropic oscillator in NC-deformed space under consideration is merely an angular momentum operator, which commutes with the rest part of Hamiltonian in the equivalent commutative space. Thus, the ground state (Gaussian) is just product of two ordinary oscillators, results into the separable states. However, for anisotropic oscillator, the deformed angular momentum operator does not commute with the rest portion of the Hamiltonian, which results into the entangled state. Thus, we can conclude that the entanglement is created via the NC-deformation of space for anisotropic oscillator.   Moreover, even in NC-space, an anisotropic oscillator also supports separable states. For instance, let us consider without loss of generality $m_1m_2=1, \hbar=1, \mu=1,\nu=1$, which provides the relation $\tilde{\omega}_2=1/\tilde{\omega}_1$ between the oscillator frequencies for which it supports separable states.
All the other frequency admits the entanglement between coordinate degrees of freedom. Since the separability of the bipartite Gaussian state for an anisotropic oscillator in NCS is satisfied for only a very special choice of anisotropic parameters, it is worth mentioning that the bipartite Gaussian state  is almost always entangled in NC-space.\\
In the next section, we elaborate our study of induced entanglement under NC deformation for  symmetric pure states.
\section{NC deformation of phase-space for symmetric pure states}  
Suppose, we have a bipartite system consisting of $A$ and $B$ in two dimensional background space (i.e., each subsystem has four dimensional phase space). The composite system has eight-dimensional phase space with co-ordinates $\hat{\xi}=(\hat{x}_{1}^{A}, \hat{x}_{2}^{A},\hat{p}_{1}^{A},\hat{p}_{2}^{A},\hat{x}_{1}^{B}, \hat{x}_{2}^{B},\hat{p}_{1}^{B},\hat{p}_{2}^{B})^T$ with the canonical commutation relations encoded in the symplectic matrix
\begin{eqnarray}
	\Omega=\mbox{Diag}(\Omega_A,\Omega_B),\; \mbox{with}\; 
	\Omega_A=\Omega_B=\left(\begin{array}{cc}
0 &\mathbb{I}_2 \\
-\mathbb{I}_2 & 0
	\end{array}\right).
\end{eqnarray} 
If, $A$ and $B$ share a composite state
\begin{equation}
	F(\xi)=\frac{1}{\pi^4\sqrt{\mbox{Det}(\Sigma)}}e^{-\xi^T\Sigma^{-1}\xi}, 
\end{equation}
then with the help of local symplectic transformation ($\mathcal{S}_{AB}=\mathcal{S}_A\oplus \mathcal{S}_B$), one can express the covariance matrix $\Sigma_{AB}$ in the  canonical form
\begin{eqnarray}\label{Sigmacanonical}
	\Sigma_{AB}=\left( \begin{array}{cc}
\sigma_{AA} & \sigma_{AB}\\
\sigma_{AB}^T & \sigma_{BB}
	\end{array}\right),
\end{eqnarray}
with
\begin{eqnarray}
	\sigma_{kk}=\bigoplus_{j=1}^{2}\sigma_{kj}\mathbb{I}_2,\; k=A,B.\\
	\sigma_{AB}=\left(\begin{array}{cc}
\mbox{Diag}(\sigma_{11},\sigma_{22}) & \mbox{Diag}(\sigma_{13},\sigma_{24})\\
\mbox{Diag}(\sigma_{13},\sigma_{24}) & \mbox{Diag}(\sigma_{33},\sigma_{44}
	\end{array}\right).
	\end{eqnarray}
Thus the composite bipartite system in  two dimensional background space can be expressed through the ten parameters 
$ (\sigma_{A1},\sigma_{A2},\sigma_{B1},\sigma_{B2},\sigma_{11},\sigma_{22},\sigma_{13},\sigma_{24},\sigma_{33},\sigma_{44}).
$
Since our aim is to focus on entanglement induced by the congruence, we restrict ourselves to symmetric pure states. In particular, we restrict ourselves to the following parameter values.
\begin{eqnarray}
\sigma_{A1}=\sigma_{A2}=\sigma_{B1}=\sigma_{B2}=b/2,\;\mbox{with}\; b>0.\\
\sigma_{11}=\sigma_{22}=-\sigma_{33}=-\sigma_{44}=nb/2,\\
\sigma_{13}=-\sigma_{24}=mb/2,\; \mbox{with}\; m,n\in \mathbb{R}.
\end{eqnarray}
The covariance matrix ~\eqref{Sigmacanonical} is rewritten in terms of the parameters $m,n,b$ as
\begin{eqnarray}
	\Sigma_{AB}=\frac{b}{2}\left(\begin{array}{cc}
\mathbb{I}_4 & \gamma^T\\
\gamma & \mathbb{I}_4
	\end{array}\right),\;
\mbox{with}\;
\gamma=\left(\begin{array}{cc}
n\mathbb{I}_2 & m\sigma_z \\
m\sigma_z & -n\mathbb{I}_2
\end{array}\right).
\end{eqnarray}
Here, we have used the following reprentation for the Pauli matrices.
\begin{eqnarray}
	\sigma_x=\left(\begin{array}{cc}
0 & 1\\
1& 0
	\end{array}\right),\; 
	\sigma_y=\left(\begin{array}{cc}
	0 & -i\\
	i& 0
\end{array}\right),\;
	\sigma_z=\left(\begin{array}{cc}
	1 & 0\\
	0& -1
\end{array}\right).
\end{eqnarray}
For simplicity let us  choose 
\begin{equation}
b=(1+R)/(1-R),\; \mbox{with}\;	R=\sqrt{m^2+n^2}.
\end{equation}
Since $b>0$, we have $R<1$. In other words, $m,n$ lies inside a unit circle, centered at the origin. Clearly, $b>1$.\\
If $\pm  \lambda_{k}\in\mathbb{R},\; (k=1,...4)$ are the eigenvalues of $2i\Omega^{-1}\Sigma_{AB}$, then the diagonal entities (symplectic eigenvalues of $\Sigma_{AB}$) $\mbox{Diag}(\sigma_z,\sigma_z,\sigma_z,\sigma_z)\mbox{Diag}(\lambda_1,-\lambda_1,\lambda_2,-\lambda_2,\lambda_3,-\lambda_3,\lambda_4,-\lambda_4)$ are the Williamson invariants. For a covariance matrix, the smallest Williamson invariant has to be greater than one, which is equivalent statement to RSUP $\Sigma_{AB}+(i/2)\Omega\ge 0$. In the present case, all the symplectic eigenvalues are same and equal to
\begin{equation}\label{RSUPspec}
	\lambda_{\Sigma,min}=(1+R)\sqrt{b}>1,\;\forall m,n
\end{equation}
Therefore $\Sigma_{AB}$ is a valid covariance matrix for a quantum system. Now, in order to determine the separability of the state let us perform the partial transposition (PT) operation on the party $B$. The PT transformation changes the sign of momentum co-ordinates of $B$:
\begin{equation}\label{PToperation}
	\xi\mapsto \xi'=\Lambda \xi,\; \mbox{with}\; \Lambda=\Lambda_A\oplus\Lambda_B,\; \Lambda_A=\mathbb{I}_4,\; \Lambda_B=\mbox{Diag}(\mathbb{I}_2,-\mathbb{I}_2).
\end{equation} 
Under PT, the covariance matrix of a separable state will be transformed into a bonafide covariance matrix. In other words,
\begin{equation}
	\Sigma'_{AB}+(i/2)\Omega\ge 0, \; \mbox{where}\; \Sigma'_{AB}=\Lambda\Sigma_{AB}\Lambda^T.
\end{equation}
The symplectic invariants of $\Sigma'_{AB}$ are 
\begin{equation}
	\lambda_{\Sigma'}=b(1\pm R).
\end{equation}
The smallest symplectic invariant ($	\lambda_{\Sigma',min}=b(1-R)= 1+R>1$) is always greater than one. Thus, we start with a separable bipartite Gaussian state. We wish to explore how Bopp's shift affects separability. Let us consider the  transformation
\begin{eqnarray}
	\hat{\tilde{\xi}}=S\hat{\xi},\; \mbox{with}\;	S=\mbox{Diag}(S_A,S_B),\;  S_A=S_B= \left( \begin{array}{cc}
		\mathbb{I}_2 & -i\frac{\theta}{2}\sigma_y \\
		i\frac{\eta}{2}\sigma_y & \mathbb{I}_2
	\end{array}\right).
\end{eqnarray}
Since $S$ is invertible we have the following constraints on the parameters
\begin{equation}
	\Delta_S=(1-\theta\eta/4)^4\neq 0,\; \Delta_{S_A}=\Delta_{S_B}=(1-\theta\eta/4)^2 \neq 0. 	
\end{equation}
Here we have used the notation $\Delta_U=\mbox{Det}(U)$.
The canonical commutation relations are encoded in the deformed symplectic matrix
\begin{eqnarray}
	\tilde{\Omega}=S\Omega S^T=\mbox{Diag}(\tilde{\Omega}_A,\tilde{\Omega}_B),
\end{eqnarray}
where
\begin{eqnarray}
	\tilde{\Omega}_A=\tilde{\Omega}_B= S_A \Omega_A S_A^T=\left(\begin{array}{cc}
		i\theta \sigma_y & \hbar_e\mathbb{I}_2\\
		-\hbar_e\mathbb{I}_2 & i\eta \sigma_y
	\end{array}\right),\; \mbox{with}\; \hbar_e=1+\theta\eta/4.
\end{eqnarray}
which corresponds to the noncommutative (NC) deformation of phase-space with the position-position NC parameter $\theta$, and momentum-momentum NC parameter $\eta$, given as
\begin{eqnarray}
	[\hat{\tilde{x}}_1^K,\hat{\tilde{x}}_2^K]=i\theta,\; [\hat{\tilde{p}}_1^K,\hat{\tilde{p}}_2^K]=i\eta,\;
	[\hat{\tilde{x}}_a^K,\hat{\tilde{p}}_b^K]=i\hbar_e\delta_{ab};\; K=A,B.
\end{eqnarray}
The non-symplectic transformation $S$ is a Darboux transformation that connects the NC space and the usual commutative space.
The covariance matrix transforms under $S$ as
\begin{equation}
	\Sigma_{AB}\mapsto \tilde{\Sigma}_{AB}=S\Sigma_{AB}S^T.
\end{equation}
The modified RSUP $\tilde{\Sigma}_{AB}+\frac{i}{2}\tilde{\Omega}\ge 0$ can be stated in terms of the spectrum of $2i\tilde{\Omega}^{-1}\tilde{\Sigma}_{AB}$. All the symplectic spectra are the same and are given by
\begin{equation}
	\lambda_{\tilde{\Sigma},min}=(1+R)\sqrt{b}>1,\;\forall m,n,\theta,\eta,
\end{equation}
which is same as ~\eqref{RSUPspec}. That means the covariance matrix $\Sigma_{AB}$ is transformed into another bonafide covariance matrix $\tilde{\Sigma}_{AB}$ under the Darboux transformation.
To envisage the separability, we perform PT operation ~\eqref{PToperation} on the party $B$. The partial transposition transforms $\tilde{\Sigma}_{AB}$ to $\tilde{\Sigma}'_{AB}=\Lambda\tilde{\Sigma}_{AB}\Lambda^T$. Four symplectic invariants for $2i\tilde{\Omega}^{-1}\tilde{\Sigma}'_{AB}$ are given by
\begin{equation}
	\lambda_{\tilde{\Sigma}'}=\left\{b\sigma_{s1},b\sigma_{s2},b\sigma_{s3},b\sigma_{s4}\right\},\mbox{with}\; \sigma_{sj}=\sqrt{\lambda_{sj}};\; j=1,..,4.
\end{equation}
Explicitly written
\begin{eqnarray}
	\lambda_{s1}= \lambda_{s10} +\frac{1}{2}\sqrt{\lambda_{s11}} +\frac{1}{2}\sqrt{\lambda_{s12} - \frac{\lambda_{s14}}{4\sqrt{\lambda_{s11}}}},\\
		\lambda_{s2} = \lambda_{s10} + \frac{1}{2}\sqrt{\lambda_{s11}}-\frac{1}{2}\sqrt{\lambda_{s12}-\frac{\lambda_{s14}}{4\sqrt{\lambda_{s11}}}} \\
			\lambda_{s3} = \lambda_{s10} - \frac{1}{2}\sqrt{\lambda_{s11}}+\frac{1}{2}\sqrt{\lambda_{s12}+\frac{\lambda_{s14}}{4\sqrt{\lambda_{s11}}}}\\
				\lambda_{s4} = \lambda_{s10} - \frac{1}{2}\sqrt{\lambda_{s11}} - \frac{1}{2}\sqrt{\lambda_{s12}+\frac{\lambda_{s14}}{4\sqrt{\lambda_{s11}}}}
\end{eqnarray}
Where
\begin{eqnarray}
	\lambda_{s10} &=& \frac{1}{\Delta_S}(\theta^2+\hbar_{e}^2)(\eta^2 + \hbar_{e}^2) + \frac{1}{\Delta_{S_A}}( \theta\eta + \hbar_e^2)R^2, \\
	\lambda_{s11} &=& \frac{\hbar_e^4}{\Delta_S^2}((\theta+\eta)^2+4\Delta_{S_A}R^2),\\
			\lambda_{s12} &=& \frac{\hbar_e^4}{4\Delta_S^2} [16\theta\eta\Delta_S (m^4+n^4)+ 
			16\Delta_{S_A}(1+\frac{1}{2}(\theta^2+\eta^2)( \theta\eta + 3\hbar_e^2)  + \\
		&&	\frac{\theta^2\eta^2}{4^4} (30 +(4+\theta\eta/4)^2))
			R^2 +  2\theta\eta\Delta_{S_A}m^2n^2  + 8(\theta+\eta)^2(\theta^2+\hbar_e^2)(\eta^2+\hbar_e^2)],\\
	\lambda_{s14} &=& -\frac{(\theta+\eta)^2 \hbar_e^4}{4\Delta_S^3}( ( \theta + \eta )^2 + 4 \Delta_{S_A}R^2). 
\end{eqnarray}
From the expressions of the singular values, it is evident that $\lambda_{s4}$ is the minimum among $\lambda_{sj}$'s.  The requirement for the separability of states in the deformed space thus turns out to be
\begin{equation}
	b\sqrt{\lambda_{s4}}\ge 1.
\end{equation}
The functional dependence of $\lambda_{s14}$ on $\theta$ and $\eta$ are similar at limiting case. In particular,
\begin{eqnarray}
	\lambda_{s14\theta}&=&\lim\limits_{\eta\to 0}	\lambda_{s14}=\chi(\theta),
	\lambda_{s14\eta}=\lim\limits_{\theta\to 0}	\lambda_{s14}=\chi(\eta),\\
	\mbox{with}\;
	\chi(\theta)&=&(1+\theta^2+R^2)-\frac{1}{2}\sqrt{\theta^2+4R^2} \nonumber \\
	&&	-\frac{1}{8} \sqrt{32(2+3\theta^2)R^2+\theta^2(32(1+\theta^2)-\sqrt{\theta^2+4R^2})}.
\end{eqnarray}
Therefore, it is sufficient to study the dependency of separability on any one of the parameters $\theta$ and $\eta$. We shall study the dependency on $\theta$ only. 
\subsection{Case-1: Separability of states depends on $\theta$}
\begin{figure}
	\includegraphics[width=8cm]{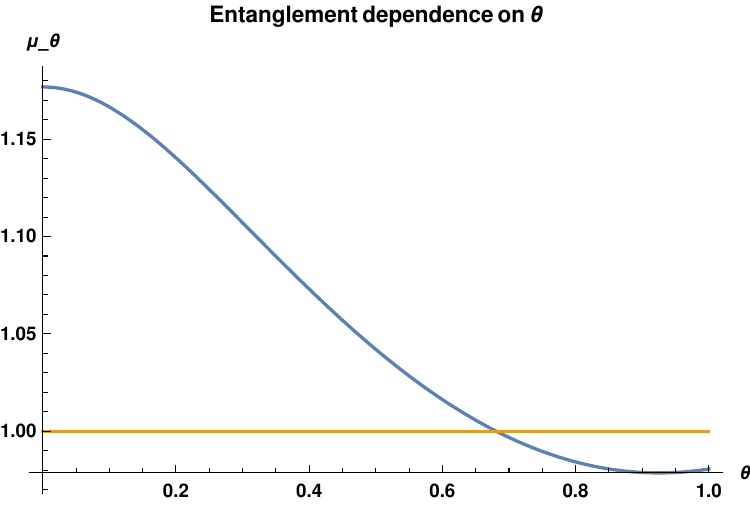}
	\caption{The blue curve indicates the value of minimum eigenvalue with respect to NC parameter $\theta\in (0,1)$. Horizontal line shows the $\mu_{-}\theta=1$ level reference. For larger value of $\theta$, the eigenvalue becomes less than one, which means the states are entangled. Here we take the parameter values $m=1/8,n=1/8$. For small values of $\theta$ states are still separable. This confirms that the entanglement is generated by  $\theta$.  }
	\label{symptransmin1}
\end{figure}
In the Figure: FIG.\ref{symptransmin1}, we plot the smallest Williamson invariant $b\sigma_{s4}$ with respect to the position-position NC-parameter $\theta\in (0,1)$. $b\sigma_{s4}\ge 1$ implies that even after PT operation, the CVM remains a bonafide CVM. The measurements of party $A$ and $B$ remain independent-the state is separable. Otherwise, something has happened that ceased the $\tilde{\Sigma}$ to be a bonafide CVM- the subsystems become correlated- the state is entangled. We fix the correlations for the original CVM by $m=n=1/8$. Figure: FIG.\ref{symptransmin1} indicates that the entanglement is generated for a large value of $\theta$. 
\subsection{Case-2: Confirmation of separability depends on $m,n$}
\begin{figure}
	\includegraphics[width=8cm]{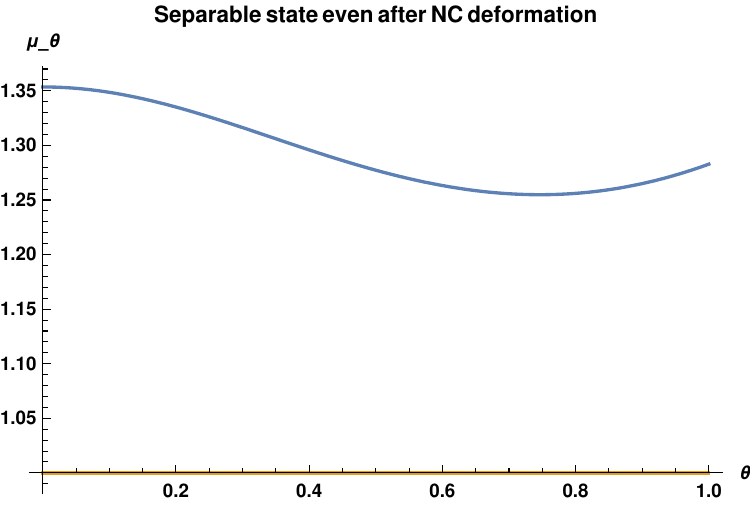}
	\caption{The blue curve indicates the value of minimum eigenvalue with respect to NC parameter $\theta\in (0,1)$. Horizontal line shows the $\mu_{-}\theta=1$ level reference. For all values of $\theta$, the smallest eigenvalue is greater than one, which means the states are separable. Here we take the parameter values $m=1/4,n=1/4$. This indicates the values of $m,n$ (correlation between off-diagonal elements) determine the separability. Separable state might map on separable state after congruence.  }
	\label{sepsattemn}
\end{figure}
Figure: FIG.\ref{sepsattemn} shows that for a specific choice of the correlations $m$ and $n$, all possible values of NC-deformation parameter $\theta$ are unable to induce the entanglement. In this instance, we have chosen $m=n=1/4$
\subsection{Case-3: Smaller the $m,n$, stronger the entanglement}
\begin{figure}
	\includegraphics[width=8cm]{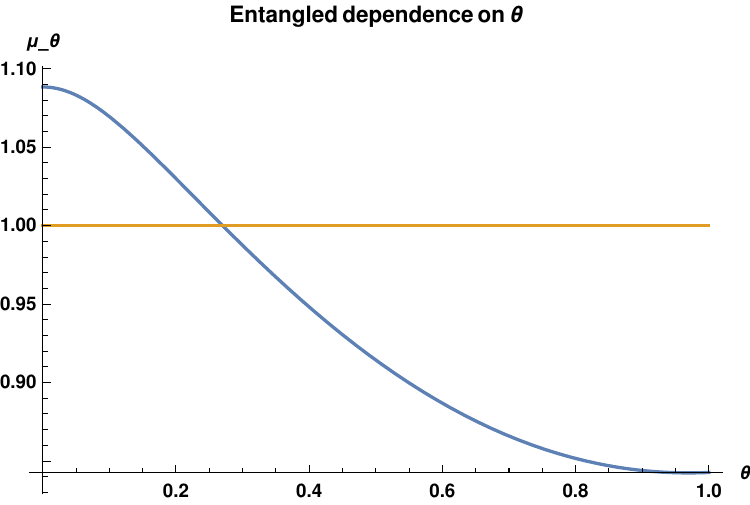}
	\caption{The blue curve indicates the value of minimum eigenvalue with respect to NC parameter $\theta\in (0,1)$. Horizontal line shows the $\mu_{-}\theta=1$ level reference.  Here we take the parameter values $m=1/16,n=1/16$. This figure indicates smaller the $m,n$, the stronger the entanglement.   }
	\label{entsatte}
\end{figure} 
Figure: FIG.\ref{entsatte} indicates that the smaller the $m,n$, the larger the allowed range of $\theta$ to generate the entanglement.
\section{Device of a gedankenexperiment: Optomechanical scheme}
Until now, in the present paper, we have discussed the entanglement generated dynamically. In particular, we devised the entanglement through Bopp's shift. In a real-life scenario, however, the two parties, Alice (A) and Bob (B), are most likely ignorant about whether the background space is commutative or noncommutative. They have to perform an experiment to determine whether the background space-time is deformed. This can be done by comparing their results with the predictions of ordinary quantum mechanics in commutative space. In this section, we discuss a gedankenexperiment through an optomechanical scheme. 
\begin{figure}
	\includegraphics[width=8cm]{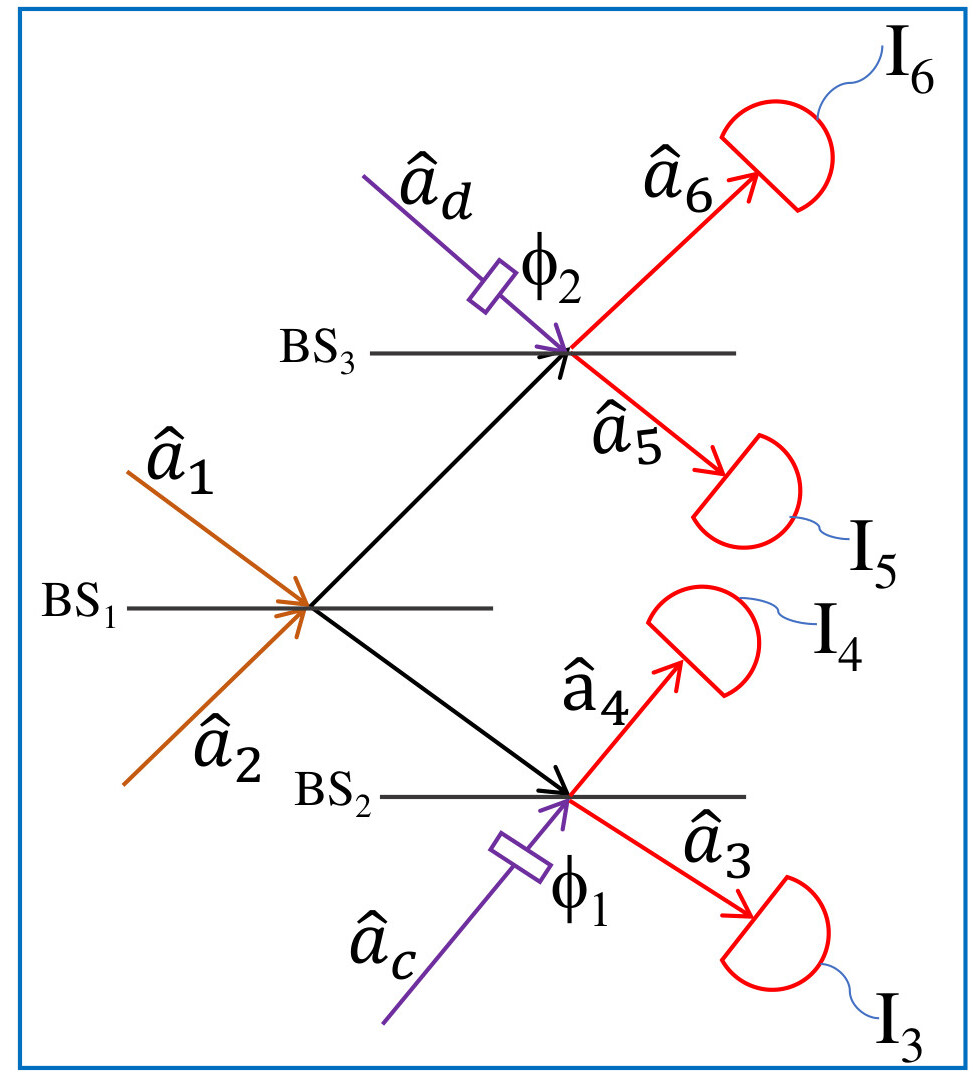}
	\caption{{\bf Determination of covariance matrix through interferometry:}    }
	\label{opto}
\end{figure}
Suppose A and B share a two-mode oscillator state determined by $\hat{a}_1$ and $\hat{a}_2$. A and B assume that background space is commutative. However, if the background state is noncommutative, then they actually share two modes of the form ~\eqref{a1a2}. We have outlined a schematic diagram for the experimental setup in Figure-FIG.\ref{opto}. Suppose the two modes $\hat{a}_1$ and $\hat{a}_2$ interfere at a 50-50 beam splitter ($BS_1$). The outputs $BS_1$ are made to interfere at another two 50-50 beam-splitters $BS_2$ and $BS_3$, along with two reference states corresponding to $\hat{a}_c$ and $\hat{a}_d$, which are phase shifted at angles $\phi_1$ and $\phi_2$, respectively, through a phase-shifter. The output of the measurement set up are represented by $\hat{a}_3$, $\hat{a}_4$, $\hat{a}_5$ and $\hat{a}_6$. In particular,
We have the following operators corresponding to the output channels
\begin{eqnarray}
	\hat{a}_3=\frac{1}{2}(\hat{a}_1-\hat{a}_2)-\frac{1}{\sqrt{2}}\hat{a}_ce^{i\phi_1} ,\\
	\hat{a}_4=\frac{1}{2}(\hat{a}_1-\hat{a}_2)+\frac{1}{\sqrt{2}}\hat{a}_ce^{i\phi_1}, \\
	\hat{a}_5=\frac{1}{2}(\hat{a}_1+\hat{a}_2)-\frac{1}{\sqrt{2}}\hat{a}_de^{i\phi_2} ,\\
	\hat{a}_6 =\frac{1}{2}(\hat{a}_1+\hat{a}_2)+\frac{1}{\sqrt{2}}\hat{a}_de^{i\phi_2},
\end{eqnarray}
where the field modes for mechanical oscillators, namely $\hat{a}_1$ and $\hat{a}_2$ are the same as those of ~\eqref{a1a2}.
The detectors generates photocurrents ($(I_3,I_4)$ and $(I_5,I_6)$) proportional to the intensities of the output modes of $BS_2$ and $BS_3$, respectively. For instance, the intensity difference at the $BS_2$ is computed as
\begin{equation}
	\hat{S}_1(\phi_1)=I_4-I_3 =\hat{a}_4^\dagger \hat{a}_4 - \hat{a}_3^\dagger \hat{a}_3,
\end{equation}
which reads explicitly as
\begin{equation}
\hat{S}_1(\phi_1)=(l_{13}\hat{x}_2 + l_{12}\hat{p}_1)\hat{x}_c^{\phi_1} + (l_{11}\hat{x}_1 + l_{14}\hat{p}_2)\hat{p}_c^{\phi_1}.
\end{equation}
We have used the notation
\begin{equation}
	\hat{x}_c^\phi=\hat{x}_c \cos\phi -\hat{p}_c \sin\phi,\; \hat{p}_c^\phi=\hat{x}_c \sin\phi +\hat{p}_c \cos\phi .
\end{equation}
The coefficients ($l_{jl},\;j=1,2;\; l=1,2,3,4$) are related to the original parameters as
\begin{equation}
	l_{jl}= k_1 \kappa_{1,l}+ (-1)^j k_2 \kappa_{2,l},
\end{equation}
where $k_j$ and $\kappa_{j,l}$ are given by \eqref{normalizationofchi}, \eqref{kappaj12} and \eqref{kappaj34}.
For unbiased reference state ($\langle \hat{x}_c\hat{p}_c\rangle =i/2$ and $\langle \hat{p}_c\hat{x}_c\rangle =-i/2$),  we get the following equations.
\begin{eqnarray}
	\langle\hat{S}_1^2(\phi_1 =0)\rangle &=& \langle \hat{x}_c^2\rangle (l_{12}^2 \langle \hat{p}_1^2\rangle + l_{13}^2 \langle \hat{x}_2^2\rangle + l_{12}l_{13} \langle \left\{ \hat{x}_2,\hat{p}_1\right\}\rangle) \nonumber \\
	&& + \langle \hat{p}_c^2\rangle (l_{11}^2 \langle \hat{x}_1^2\rangle + l_{14}^2 \langle \hat{p}_2^2\rangle + l_{11}l_{14} \langle \left\{ \hat{x}_1,\hat{p}_2\right\}\rangle) +\frac{1}{2}(l_{11}l_{12}-l_{13}l_{14}).
\end{eqnarray}
\begin{eqnarray}
\langle\hat{S}_1^2(\phi_1 =\pi/2)\rangle &=& \langle \hat{x}_c^2\rangle (l_{11}^2 \langle \hat{x}_1^2\rangle + l_{14}^2 \langle \hat{p}_2^2\rangle + l_{11}l_{14} \langle \left\{ \hat{x}_1,\hat{p}_2\right\}\rangle) \nonumber \\
&& + \langle \hat{p}_c^2\rangle (l_{12}^2 \langle \hat{p}_1^2\rangle + l_{13}^2 \langle \hat{x}_2^2\rangle + l_{12}l_{13} \langle \left\{ \hat{x}_2,\hat{p}_1\right\}\rangle) +\frac{1}{2}(l_{11}l_{12}-l_{13}l_{14}).
\end{eqnarray}
\begin{eqnarray}
	\langle\hat{S}_1^2(\phi_1 =\pi/4)\rangle &=& \frac{1}{2} (\langle \hat{x}_c^2\rangle + \langle \hat{p}_c^2\rangle) (l_{11}^2\langle \hat{x}_1^2\rangle + l_{12}^2\langle \hat{p}_1^2\rangle + l_{13}^2\langle \hat{x}_2^2\rangle + l_{14}^2\langle \hat{p}_2^2\rangle \nonumber \\ 
	&& + l_{11}l_{14} \langle \left\{\hat{x}_1, \hat{p}_2\right\} \rangle 
	+ l_{12}l_{13} \langle \left\{\hat{x}_2, \hat{p}_1\right\} \rangle) + \frac{1}{2} (\langle \hat{x}_c^2\rangle - \langle \hat{p}_c^2\rangle) (  l_{11}l_{13} \langle \left\{\hat{x}_1, \hat{x}_2\right\}\rangle \nonumber \\ 
&&	+  l_{13}l_{14} \langle \left\{\hat{x}_2, \hat{p}_2\right\} \rangle 
	+  l_{11}l_{12} \langle \left\{\hat{x}_1, \hat{p}_1\right\} \rangle +  l_{12}l_{14} \langle \left\{\hat{p}_1, \hat{p}_2\right\} \rangle).
\end{eqnarray}
Similarly
\begin{equation}
	\hat{S}_1(\phi_2)=I_6-I_5 =\hat{a}_6^\dagger \hat{a}_6 - \hat{a}_5^\dagger \hat{a}_5,
\end{equation}
which reads explicitly as
\begin{equation}
	\hat{S}_1(\phi_2)=(l_{23}\hat{x}_2 + l_{22}\hat{p}_1)\hat{x}_d^{\phi_2} + (l_{21}\hat{x}_1 + l_{24}\hat{p}_2)\hat{p}_d^{\phi_2}.
\end{equation}
\begin{eqnarray}
	\langle\hat{S}_1^2(\phi_2 =0)\rangle &=& \langle \hat{x}_d^2\rangle (l_{22}^2 \langle \hat{p}_1^2\rangle + l_{23}^2 \langle \hat{x}_2^2\rangle + l_{22}l_{23} \langle \left\{ \hat{x}_2,\hat{p}_1\right\}\rangle) \nonumber \\
	&& + \langle \hat{p}_d^2\rangle (l_{21}^2 \langle \hat{x}_1^2\rangle + l_{24}^2 \langle \hat{p}_2^2\rangle + l_{21}l_{24} \langle \left\{ \hat{x}_1,\hat{p}_2\right\}\rangle) +\frac{1}{2}(l_{21}l_{22}-l_{23}l_{24}).
\end{eqnarray}
\begin{eqnarray}
	\langle\hat{S}_1^2(\phi_2 =\pi/2)\rangle &=& \langle \hat{x}_d^2\rangle (l_{21}^2 \langle \hat{x}_1^2\rangle + l_{24}^2 \langle \hat{p}_2^2\rangle + l_{21}l_{24} \langle \left\{ \hat{x}_1,\hat{p}_2\right\}\rangle) \nonumber \\
	&& + \langle \hat{p}_d^2\rangle (l_{22}^2 \langle \hat{p}_1^2\rangle + l_{23}^2 \langle \hat{x}_2^2\rangle + l_{22}l_{23} \langle \left\{ \hat{x}_2,\hat{p}_1\right\}\rangle) +\frac{1}{2}(l_{21}l_{22}-l_{23}l_{24}).
\end{eqnarray}
\begin{eqnarray}
	\langle\hat{S}_1^2(\phi_2 =\pi/4)\rangle &=& \frac{1}{2} (\langle \hat{x}_d^2\rangle + \langle \hat{p}_d^2\rangle) (l_{21}^2\langle \hat{x}_1^2\rangle + l_{22}^2\langle \hat{p}_1^2\rangle + l_{23}^2\langle \hat{x}_2^2\rangle + l_{24}^2\langle \hat{p}_2^2\rangle \nonumber \\ 
	&& + l_{21}l_{24} \langle \left\{\hat{x}_1, \hat{p}_2\right\} \rangle 
	+ l_{22}l_{23} \langle \left\{\hat{x}_2, \hat{p}_1\right\} \rangle) + \frac{1}{2} (\langle \hat{x}_d^2\rangle - \langle \hat{p}_d^2\rangle) (  l_{21}l_{23} \langle \left\{\hat{x}_1, \hat{x}_2\right\}\rangle \nonumber \\ 
	&&	+  l_{23}l_{24} \langle \left\{\hat{x}_2, \hat{p}_2\right\} \rangle 
	+  l_{21}l_{22} \langle \left\{\hat{x}_1, \hat{p}_1\right\} \rangle +  l_{22}l_{24} \langle \left\{\hat{p}_1, \hat{p}_2\right\} \rangle).
\end{eqnarray}
Joint measurement on the two output modes of beam-splitters $BS_2$ and $BS_3$ is given by
\begin{eqnarray}
\hat{S}_3(\phi_1,\phi_2)=	i(\hat{a}_6^\dagger \hat{a}_2-\hat{a}_3^\dagger \hat{a}_6) = \frac{1}{4}(l_{12}l_{21}-l_{11}l_{22})\{\hat{x}_1,\hat{p}_1\} + \frac{1}{4}(l_{13}l_{24}-l_{14}l_{23})\{\hat{x}_2,\hat{p}_2\} \nonumber \\
	 + \frac{1}{4}(l_{12}l_{24}-l_{14}l_{22})\{\hat{p}_1,\hat{p}_2\} +
	\frac{1}{4}(l_{13}l_{21}-l_{11}l_{23})\{\hat{x}_1,\hat{x}_2\} 
	 - \frac{1}{2}(l_{11}\hat{x}_1+l_{14}\hat{p}_2)\hat{x}_d^{\phi_2} \nonumber \\
	 - \frac{1}{2}(l_{21}\hat{x}_1+l_{24}\hat{p}_2)\hat{x}_c^{\phi_1}  +\frac{1}{2}(l_{23}\hat{x}_2+l_{22}\hat{p}_1)\hat{p}_c^{\phi_1} + \frac{1}{2}(l_{13}\hat{x}_2+l_{12}\hat{p}_1)\hat{p}_d^{\phi_2} \nonumber \\
	 + (\{\hat{x}_c,\hat{x}_d\}+\{\hat{p}_c,\hat{p}_d\})\sin(\phi_1-\phi_2) + (\{\hat{x}_d,\hat{p}_c\}- \{\hat{x}_c,\hat{p}_d\})\cos(\phi_1-\phi_2).
\end{eqnarray}
One can measure the expectation values of $\hat{S}_3$ for various angles similar to $\hat{S}_1^2$ and write down another linear equation. Solving all those linear equations, one can collect the data of expectation values of all the quadratures and explicitly write down the covariance matrix. This covariance matrix will show the entanglement property through the Peres-Horodecki criteria (Simon's condition) as stated in ~\eqref{separabilityc}. However, Alice and Bob thought that they had shared a separable state in commutative space. This entanglement is generated through the noncommutative parameters ($\theta,\eta$) of the background noncommutative space.
\section{Conclusions}
In the present work, we revisited the problem of entanglement in Gaussian states induced by deformations of phase space. The relevance of considering Gaussian states is twofold: first, Gaussian states are the most commonly experimentally used continuous-variable states. Second, the Gaussian quantum
states are represented in the phase space picture of quantum mechanics as a proper probability distribution function, coming from Wigner functions in the phase space. 
The positive partial transpose criterion for entanglement separability of bipartite Gaussian states is extended for a general class of Bopp's shift, which falls under the category of $Gl(2n,\mathbb{R})$.  In particular, we have considered both the position-position and momentum-momentum NC-deformation, with deformation parameters $\theta$ and $\eta$ respectively. It turns out that $\theta$ and $\eta$ induce the entanglement. We have directly applied the formalism of phase-space deformation through Bopp's shift for an anisotropic two-dimensional harmonic oscillator. Peres-Horodecki separability condition leads to a constraint equation that consists of NC parameters and parameter values of oscillators, for which the bipartite state of the oscillator is entangled or separable. It turns out that the bipartite Gaussian state is almost always entangled in NC-space. We have extended our study for a symmetric bipartite pure state, for which the entanglement generation through NC parameters is explicitly shown graphically.  We have shown that the effects of $\theta$ and $\eta$ are mathematically identical, which leads to the fact that the study of separability for any one parameter (we did with $\theta$) from $\theta$ and $\eta$ is sufficient for quantitative analysis. In particular, we have illustrated the NC-parameter dependence on entanglement generation through the effect of $\theta$ in the smallest Williamson invariant with a graphical representation. As the entanglement of Gaussian states has been shown to be created by both configuration and/or momentum variables, the one-particle sector of theories like string theory and quantum theory can exhibit the effects discussed in this work. This possibility opens a completely new strategy to test these theories.\\
In addition to the purely abstract study of induced entanglement for NC deformation of phase-space, we have outlined a gedankenexperiment to test the theory under consideration. The experimental design is important due to the fact that, in a real-life situation,  the two parties, Alice (A) and Bob (B), are most likely ignorant about whether the background space is commutative or noncommutative. However, we have outlined a simple experimental setup through interferometry.  Through interferometry, the phonon modes of two-dimensional oscillators interfere with reference light beams and produce output photocurrent in detectors. We have demonstrated how the average photocurrents are related to the expectation values of quadratures. In other words, one can simply measure the output photocurrent and compute the covariance matrix elements. Through the covariance matrix, one can apply separability criteria and determine whether the state is entangled.  If the shared oscillator state, which A and B had believed to be separable, is found to be entangled after photocurrent measurement, then one can conclude that something is mediated to create the entanglement. If no other effect is present, then it is likely that the background space-time is noncommutative. Since no classical intermediate entity can cause the entanglement, it can thus be concluded that the space-time noncommutativity is a signature of the quantum nature of gravity.
\section{Data Availability Statement}
The manuscript has no associated data.
\section{Conflict of interests}
All the authors declare that there is no conflict of interest from funding agency or any other means whatsover.  
\section{Acknowledgement}
S. Nandi and S. Maity are grateful to ANRF (Formerly SERB), Govt. of India, for fellowship support through project grant no  EEQ/2023/000784. P. Patra is grateful to ANRF (Formerly SERB), Govt. of India, for financial support through project grant No. EEQ/2023/000784.\\
We are grateful to the anonymous reviewer for fruitful suggestions, which made this manuscript in its present form.

\end{document}